\documentclass[conference]{IEEEtran}
\IEEEoverridecommandlockouts
\usepackage{cite}
\usepackage{amsmath,amssymb,amsfonts}
\usepackage{graphicx,subcaption}
\usepackage{algorithm, algorithmic}
\usepackage{textcomp}
\usepackage{xcolor}
\usepackage{epsfig}
\usepackage{epstopdf}
\usepackage{balance}
\usepackage{lipsum}  
\usepackage{tabularray}
\usepackage{stfloats}
\usepackage{lipsum}
\setlipsum{%
  par-before = \begingroup\color{gray},
  par-after = \endgroup
}

\def\BibTeX{{\rm B\kern-.05em{\sc i\kern-.025em b}\kern-.08em
    T\kern-.1667em\lower.7ex\hbox{E}\kern-.125emX}}
\begin{document}

\title{Rapid CNN-Assisted Iterative RIS Element Configuration
}

\author{\IEEEauthorblockN{
Samed Keşir\textsuperscript{$\ast$,$\circ$},
M. Yaser Yağan\textsuperscript{$\ast$,$\circ$},
İbrahim Hökelek\textsuperscript{$\ast$},
Ali Emre Pusane\textsuperscript{$\circ$},
Ali Görçin\textsuperscript{$\ast$}
}
	\IEEEauthorblockA{
	\textsuperscript{$\ast$} Communications and Signal Processing Research (HISAR) Lab., TUBITAK BILGEM, Kocaeli, Turkey\\
    \textsuperscript{$\circ$} Department of Electrical and Electronics Engineering, Boğaziçi University, Istanbul, Turkey} 
		Email:
		\{samed.kesir,yaser.yagan,ibrahim.hokelek\}@tubitak.gov.tr,\\
		ali.pusane@boun.edu.tr, ali.gorcin@tubitak.gov.tr
		\vspace{-3ex}}

\IEEEoverridecommandlockouts
\IEEEpubid{\makebox[\columnwidth]{979-8-3503-3559-0/23\$31.00 ~\copyright2023 IEEE  \hfill} \hspace{\columnsep}\makebox[\columnwidth]{ }}
\maketitle
\IEEEpubidadjcol
\begin{abstract}
Reconfigurable Intelligent Surfaces (RISs) are becoming one of the fundamental building blocks of next-generation wireless communication systems. To that end, RIS phase configuration optimization is an important issue, where finding the most suitable configuration becomes a challenging and resource-consuming task, especially as the number of RIS elements increases. Since exhaustive search is not practical, iterative algorithms are utilized to determine the RIS configuration by sequentially considering all RIS elements, where the best-performing phase shift configuration is obtained for each element. However, each configuration attempt requires receiver performance feedback, leading to higher delay and signaling overhead. Thus, in this paper, a convolutional neural network (CNN) based solution is formulated to rapidly find the phase configurations of the RIS elements. The simulation results for a RIS with $40\times40$ elements imply that the proposed algorithm reduces the number of steps dramatically e.g., from 3200 to 160 for the particular setup. Furthermore, such improvement in complexity is achieved with a slight degradation in performance.



\end{abstract}

\begin{IEEEkeywords}
reconfigurable intelligent surface, convolutional neural network. 
\end{IEEEkeywords}

\section{Introduction}

Next-generation communication networks are anticipated to massively digitize societal and industrial processes through innovative applications such as massive twinning, holographic telepresence, internet of senses, and autonomous vehicle communications. Reconfigurable Intelligent Surfaces (RISs) arise as a key technology for reliable communications under harsh wireless channel conditions. A RIS consisting of a large number of passive reflecting elements \cite{lis} allows collaboratively conveying the reflected signals towards any desired location using low-cost components such as PN diodes \cite{RISgeneral}, where each element can have ON and OFF states resulting in a reflected wave with $180^\circ$ and $0^\circ$ phase shift, respectively. More complex structures have also been proposed, where each element can have more than two states with configurable phase shifts \cite{2bitRIS,3bitRIS}.


In addition to low cost and low complexity RIS design and fabrication, defining use cases and developing algorithms for rapid RIS element configuration is one of the two main trending topics of research. As the latter is studied in this paper, a short review of RIS configuration approaches is therefore provided in this section, and in this context the works can be divided into two main groups, where the RIS configuration is performed with or without channel state information (CSI). As an example of CSI-based approaches, the authors propose a centralized algorithm based on a semi-definite relaxation (SDR) technique \cite{csi_based}. While this method determines the optimum configuration of RIS directly by a closed-form solution, estimating the CSI information at RIS incurs excessive channel estimation and signalling overhead. Therefore, a low-complexity distributed algorithm achieving near-optimal performance is proposed for practical implementation such that the access point and the RIS independently perform optimization in an alternating manner until the convergence is reached \cite{csi_based}. In \cite{sfp_relay_comp}, the authors consider an outdoor cellular network where a multiple-antenna base station reaches mobile users through a RIS; two algorithms are proposed, first employs gradient descent to obtain RIS phase coefficients while second employs sequential fractional programming to extract them. Fractional programming is utilized for optimal transmit power allocation for both algorithms. Deep learning-based algorithms for determining the configuration of the RIS elements have recently gained momentum. For example, deep reinforcement learning (DRL) is utilized for joint optimization of transmitter beamforming and RIS configuration in \cite{drl_based,drl_2}. In addition, \cite{federated} employs federated learning to perform the RIS configuration optimization for a multi-user scenario.



Considering the challenges of CSI-based approaches, optimizing the RIS configuration based on the receiver feedback is another widely used method in the literature.
In \cite{genetic_ris}, a genetic algorithm (GA) is proposed to solve the phase shifts optimization problem for a multi-pair communication system, where the achievable sum rate feedback is utilized as the fitness function.  Numerical results illustrate that the GA provides almost the same performance as the exhaustive search when the number of antenna elements in the RIS is relatively small. In another practical study \cite{iterative}, an iterative search algorithm is employed by measuring the transmission amplitude of the vector network analyzer (VNA) for all possible phase delay states of each antenna element. The state of the elements maximizing the transmission amplitude is selected as the RIS configuration. A similar iterative algorithm is also employed in \cite{coverage_enhancement} to experimentally study the impact of grouping the RIS elements, where each iteration considers a group of RIS elements rather than a single element. In \cite{MNtoM+N}, RIS is configured by grouping the elements in horizontal and vertical dimensions to significantly decrease the total number of iterations. An iterative algorithm is utilized to find a suitable RIS configuration that provides a physical layer security in \cite{pls_meas}. Simulation results are verified with the measurement experiments demonstrating that a RIS can effectively increase the secrecy capacity between intended and unintended users.

The iterative method (IM) can provide near-optimal solutions \cite{iterative}, where the state of each RIS element is altered one by one and the performance feedback obtained from the receiver is utilized to determine the best configuration for each element. However, when the number of RIS elements is large, the IM becomes costly and inefficient. In this paper, a Convolutional Neural Network (CNN) based method is proposed to rapidly find the phase configurations of RIS elements. In the proposed method a CNN model is utilized to enhance the performance of group based iterative method (G-IM) which significantly reduces the number of iterations in the IM. For a RIS composed of $N$ vertical and $M$ horizontal elements, conventional IM requires $N\times M$ iterations while the CNN-assisted G-IM (CNN-G-IM) performs only $N+M$ iterations by utilizing a trained CNN model. The idea of employing CNN for generating a final RIS configuration image from two initial RIS configuration images is presented in \cite{deep_neural}. Apart from \cite{deep_neural}, the CNN-G-IM utilizes the G-IM for finding the best horizontal and vertical RIS configurations. The trained CNN model provides the final RIS configuration by utilizing these two configurations as its inputs.  The simulation results for a RIS with $40\times40$ elements indicate that the CNN-G-IM reduces the number of steps from 3200 to 160. This significant improvement in complexity comes with only a slight degradation in performance.

The remainder of this paper is organized as follows. Section \ref{sec:SystemModel} describes the system and channel models for a RIS-assisted wireless communication. The proposed CNN-assisted group based iterative algorithm is presented in Section III. Section IV reports the simulation results. Finally, Section V concludes the paper.

\section{System and Channel Models}\label{sec:SystemModel}

\begin{figure}[t]
    \includegraphics[width=1.1\linewidth]{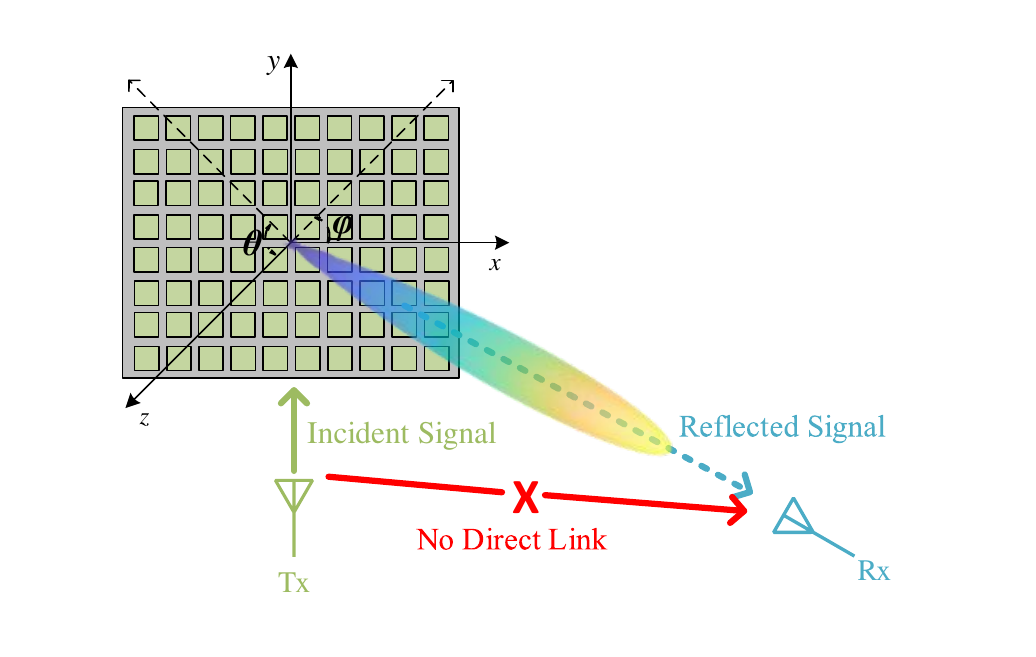}
    \caption{An RIS-assisted wireless communication system.} 
    \vspace{-10pt}
    \label{fig:system_schema}
\end{figure}

RISs are expected to play an important role in 6G by enabling new use cases. For example, RISs can form an alternative link through their controllable reflections when there is no direct link between a transmitter and a receiver. This is particularly a serious challenge at mmWave frequencies due to high path loss and blockage issues. An RIS-assisted wireless communication system is shown in Fig. \ref{fig:system_schema}, where a passive RIS is assumed to be located on the $xy$-plane and faces the $+z$ direction. The transmitter is on the opposite side and propagating in the $-z$ direction. $\theta$ and $\phi$ represent the elevation and azimuth angles according to the $xy$-plane, respectively. The passive reflecting elements of the RIS, represented by the small squares, are configurable such that each element reflects incoming signals with a phase delay of either $0^\circ$ or $180^\circ$. The transmitter (Tx) and receiver (Rx) are located in such a way that there is no direct link between them, and they only communicate through the reflected signals from the RIS. 

A passive RIS with $M$ by $N$ elements on its horizontal and vertical axes, respectively, is considered to develop an analytical model for the RIS-assisted wireless communication channel in Fig. \ref{fig:system_schema}. 
The scattering electric field of the RIS as a function of elevation and azimuth angles $(\vartheta, \varphi)$ can be obtained by the superposition of single elements' fields as \cite{yang2016programmable}
\begin{equation}
\label{eq:e-field}
\begin{aligned}
E(\vartheta, \varphi)= & cos(\vartheta) \sum_{m=0}^{M-1} \sum_{n=0}^{N-1} A_{mn} e^{j \alpha_{mn}}   cos\left(\vartheta_{mn}\right) \Gamma_{mn} e^{j \phi_{mn}} \\
& \times e^{j k_{0}\left(m d_{x} \sin \vartheta \cos \varphi+n d_{y} \sin \vartheta \sin \varphi\right)},
\end{aligned}
\end{equation}
where $cos(.)$ represents the radiation pattern of each individual antenna element in the array, and it is used for both incident and reflected waves. $A_{mn} $ and $ \alpha_{mn}$ are the illuminating amplitude and phase of the signal radiating on the RIS element ($m,n$). $\vartheta_{mn}$ and $\varphi_{mn}$ denote the elevation and azimuth angles of the transmitter. The term $e^{\phi_{m n}}$ describes the reflection phase contribution. Finally, the last term describes the array steering factor with $k_0$ presenting the free-space wave number, $d_x$ and $d_y$ representing the distances between adjacent reflecting elements in $x$ and $y$ directions, respectively. 

The channel between the transmitter and the receiver through the RIS is decomposed into two parts: $h$ from the transmitter to the RIS and $g$ from the RIS to the receiver, respectively. Since the transmitter is located in the near-field of the RIS, the radiation pattern of each element for the incident signal is represented with $\cos \vartheta_{mn}$. However, for the reflected signal, the radiation pattern of the element is represented by $\cos \vartheta_{rx}$ since the receiver is in the far field of the RIS. The channel components can be modeled using (\ref{eq:e-field}) as 
\begin{equation}
\label{eq:channels}
\begin{aligned}
h_{mn} = & L^{tx}_{mn}\cos \vartheta_{mn},\\
g_{mn} = & L^{rx}_{mn} \cos \vartheta_{rx}\\
&\times e^{j k_0\left(m d_x \sin \vartheta_{rx} \cos\varphi_{rx}+n d_y \sin \vartheta_{rx} \sin \varphi_{rx}\right)}, \\
\end{aligned}
\end{equation}
where $L^{tx}_{mn}$ and $L^{rx}_{mn}$ represent the path losses taken as free space path losses \cite{pathloss} for the $h_{mn}$ and $g_{mn}$ channels, respectively. $\vartheta_{rx}$ and $\varphi_{rx}$ are the elevation and azimuth angles of the receiver, respectively.  

Consequently, from a communication perspective, the received baseband signal $y_{rx}[k]$ can be expressed in terms of the transmitted signal $x[k]$ as
\begin{equation}
\label{eq:rec_signal}
\begin{aligned}
y_{rx}[k]=&  \sum_{m=0}^{M-1} \sum_{n=0}^{N-1} h_{mn} e^{j \phi_{mn}}  g_{mn} x[k] + n[k]
\end{aligned}
\end{equation}
with $n[k]$ being the additive noise at the receiver. Then, the received power can be estimated as 

\begin{equation}
    P_{rx}=10\log_{10} \left(\frac{1}{K}\sum_{k=1}^K y_{rx}[k]y_{rx}^*[k]\right),
\end{equation}
where $(.)^*$ is the complex conjugation operation. Assuming that $n[k]$ is the white Gaussian noise component, the received power of the receiver can be increased by optimizing the RIS as 


\begin{equation}
\label{eq:max}
\begin{aligned}
\mathbf{\Theta}^*=\arg &\max_{\phi_{mn}, \forall  mn} \left| \sum_{m=0}^{M-1} \sum_{n=0}^{N-1} h_{mn} e^{j \phi_{mn}}  g_{mn}\right|\\
&\text{ s.t. } \phi_{mn} \in\left\{0^{\circ}, 180^{\circ}\right\}, \quad \forall  mn ,
\end{aligned}
\end{equation}
where $\mathbf{\Theta}^*$ is the set of optimum phases of the RIS elements.


\section{CNN-ASSISTED ITERATIVE ALGORITHM}

\begin{figure}[t]
    \includegraphics[width=1\linewidth]{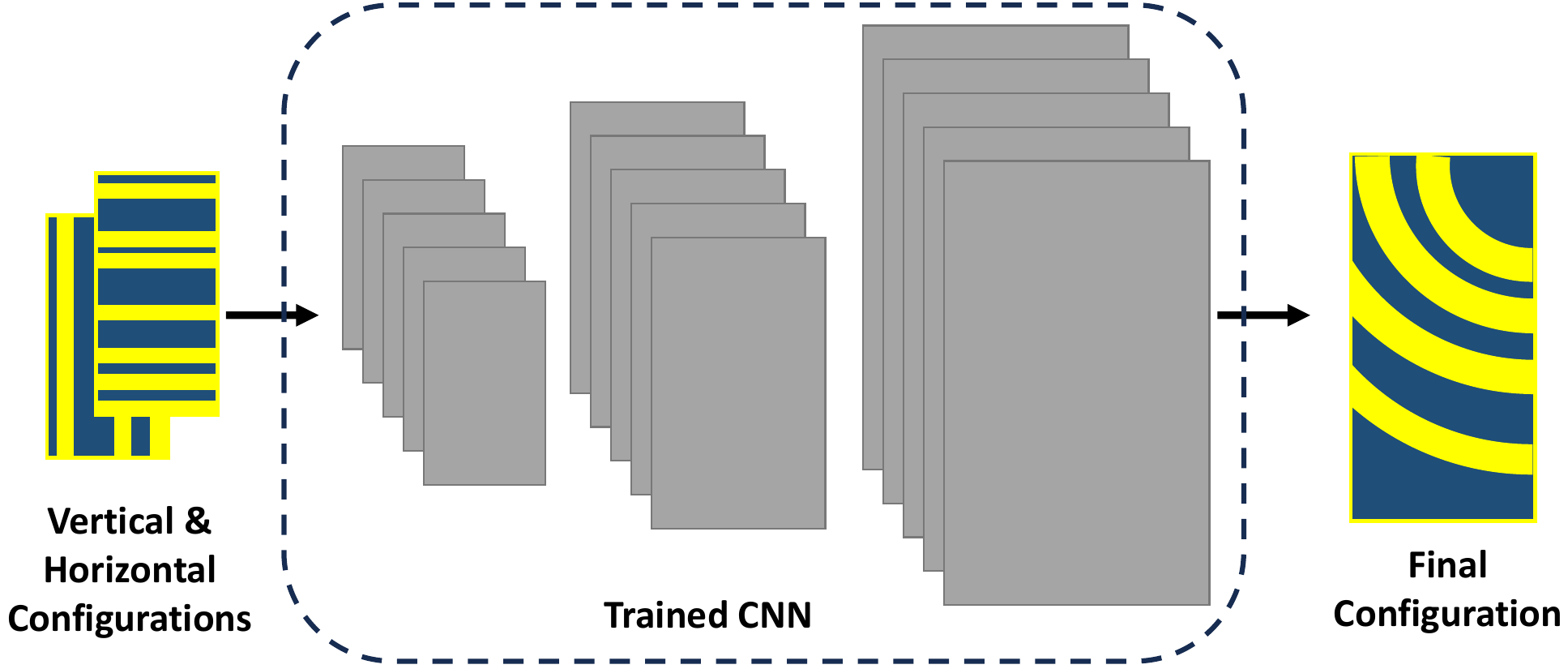}
    \caption{The proposed CNN-based RIS optimization algorithm.} 
    \vspace{-10pt}
    \label{fig:algorithm_schema}
\end{figure}

CNN is a type of deep learning algorithm commonly used in image and video processing applications \cite{cnn_im_proc,cnn_im_proc_2}. Unlike traditional neural networks, CNNs are specifically designed to process data with a grid-like topology, such as images, by utilizing a series of convolutional layers to extract relevant features from the input data. Generally, this feature extraction process is followed by one or more fully connected layers, which use the extracted features to perform classification or regression tasks. However, in this study, a CNN is not utilized to perform classification or regression but to learn the relationship between the input images and the output image in order to generate the RIS configuration. Fig. \ref{fig:algorithm_schema} shows the proposed CNN model, where horizontal and vertical configurations of a specific receiver are utilized to generate its final RIS configuration. The proposed method can rapidly find a suitable configuration which otherwise is a challenging and resource-consuming task, especially as the number of RIS elements increases.

\begin{figure*}[!t]

\begin{minipage}[h]{0.32\textwidth}
    \centering
    \includegraphics[width=1\textwidth]{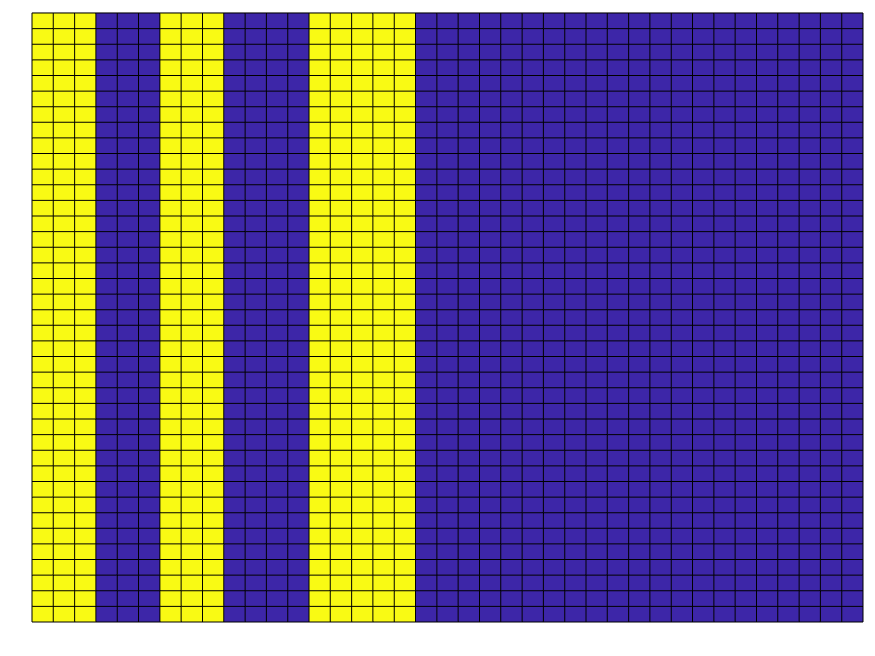}
    \subcaption{}
\end{minipage}
\begin{minipage}[h]{0.32\textwidth}
    \centering
    \includegraphics[width=1\textwidth]{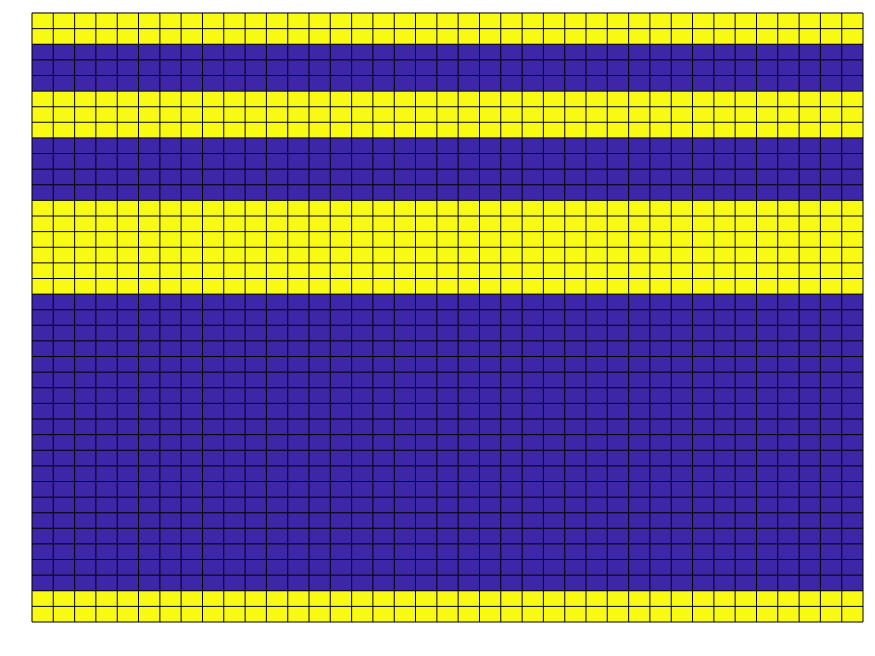}
    \subcaption{}
\end{minipage}
\begin{minipage}[h]{0.32\textwidth}
    \centering
    \includegraphics[width=1\textwidth]{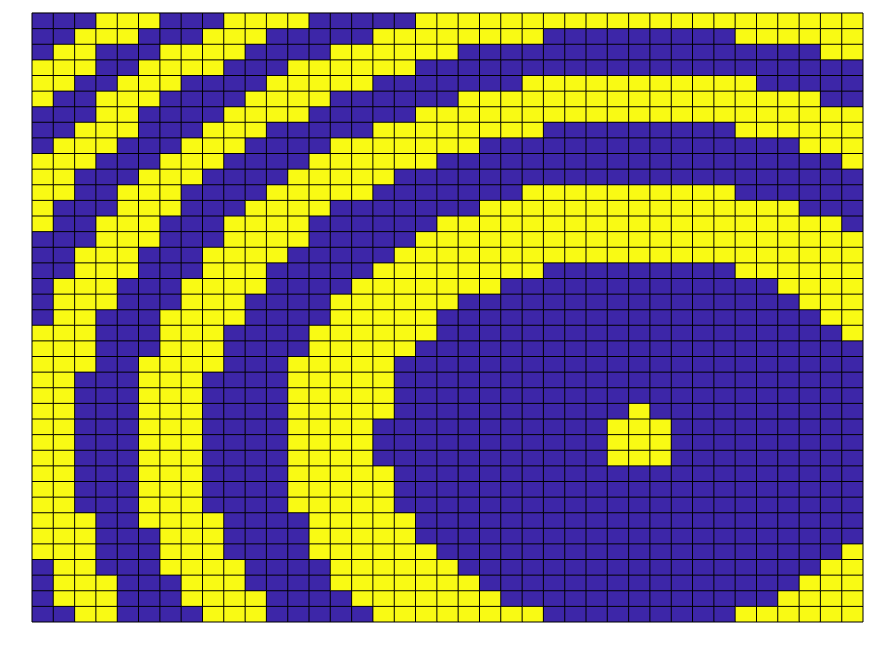}
    \subcaption{}
\end{minipage}

\begin{minipage}[h]{0.32\textwidth}
    \centering
    \includegraphics[width=1\textwidth]{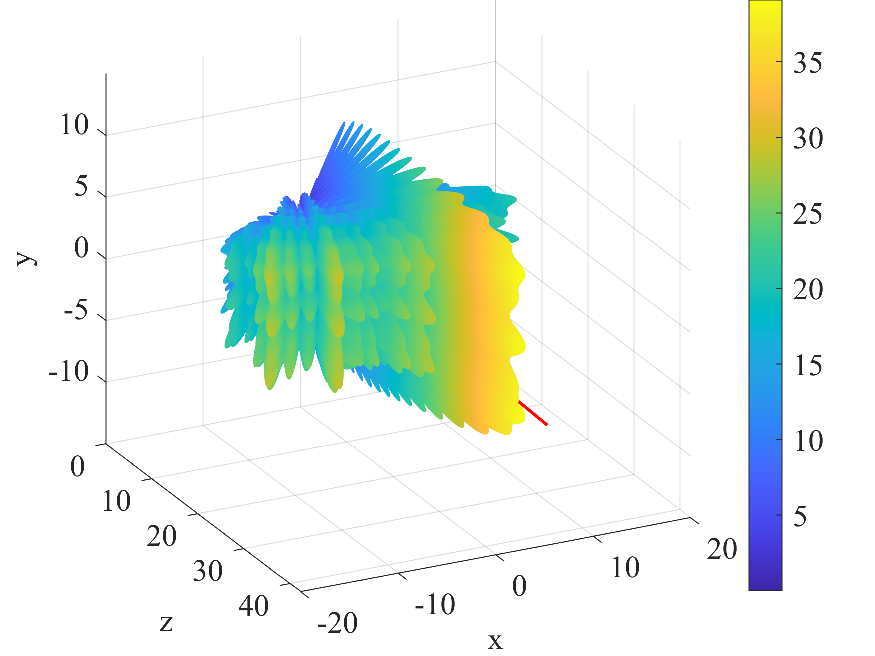}
    \subcaption{}
\end{minipage}
\begin{minipage}[h]{0.32\textwidth}
    \centering
    \includegraphics[width=1\textwidth]{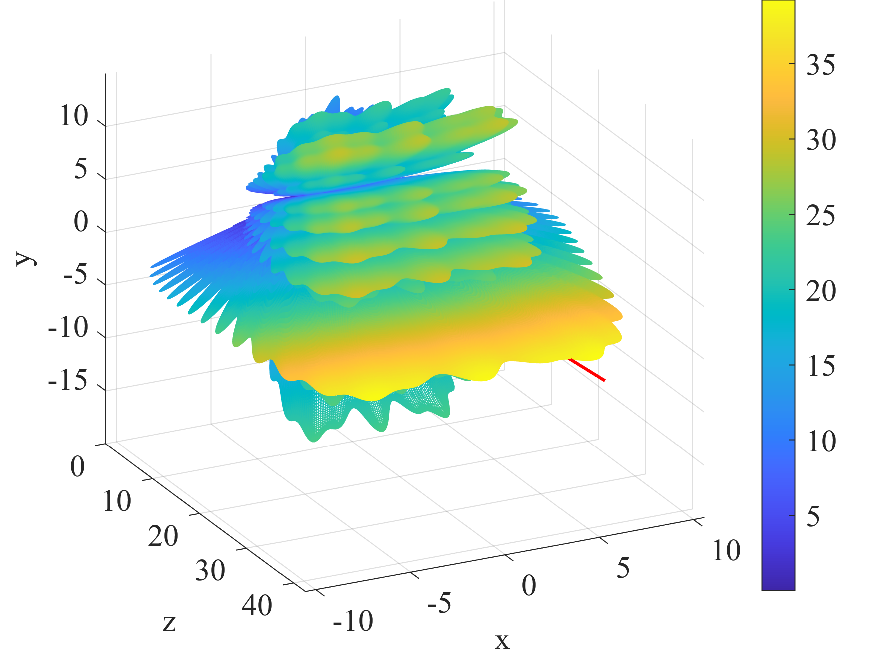}
    \subcaption{}
\end{minipage}
\begin{minipage}[h]{0.32\textwidth}
    \centering
    \includegraphics[width=1\textwidth]{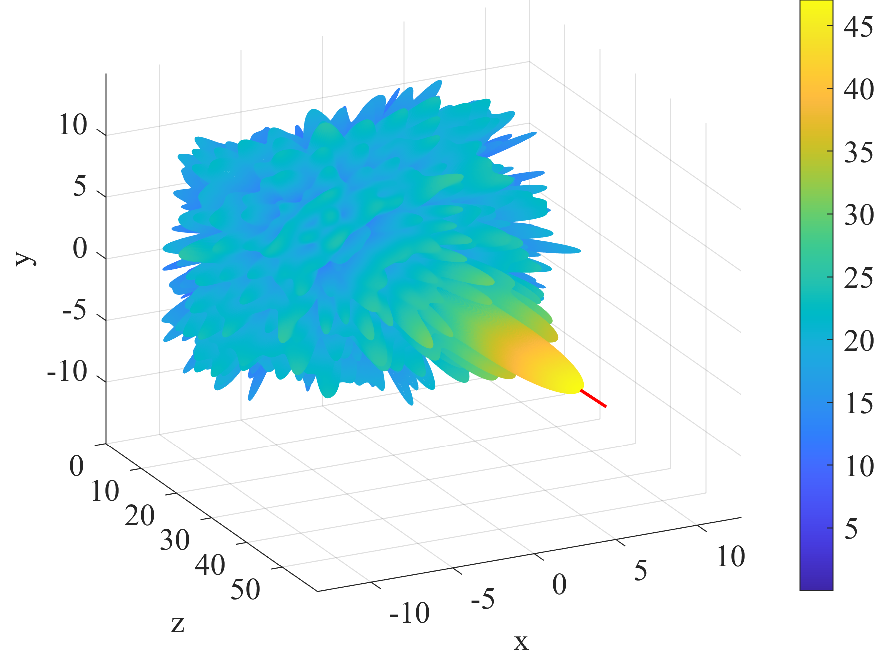}
    \subcaption{}
\end{minipage}

\caption{RIS configurations found by searching the optimum for (a) vertical beam planes, (b) horizontal beam planes, and (c) all beam planes. The radiation patterns of these configurations are given in (d), (e), and (f), respectively.}
\label{fig:rad_pattern} 
\end{figure*}

A pencil-like beam pointing in a given direction ($\vartheta_{0},\varphi_{0}$) can be obtained as the intersection of a vertical and a horizontal hyperplane in the beamspace. By examining (\ref{eq:e-field}), it can be observed that configuring the RIS elements as horizontal and vertical groups results in a vertical and horizontal beam plane, respectively. The configuration resulting from horizontally grouped elements is referred to as horizontal configuration, while the other is named vertical configuration. By grouping the elements as stripes, it is possible to configure all the elements on the same line (with the same $m$ or $n$ index) with the same phase delay at each iteration. For example, Figs. \ref{fig:rad_pattern} (a) and (b) demonstrate horizontal and vertical RIS configurations for a receiver located at an arbitrary position in Fig. \ref{fig:system_schema}. The resulting radiation patterns are shown in Figs. \ref{fig:rad_pattern} (d) and (e), respectively. At the training stage of the CNN model, these horizontal and vertical configurations are provided to the CNN model along with the reference RIS configuration as a label depicted in Fig. \ref{fig:rad_pattern} (c). Note that the reference RIS configuration is obtained using an iterative algorithm with $M\times N \times P$ iterations \cite{coverage_enhancement}. The resulting radiation pattern of the reference configuration is shown in Fig. \ref{fig:rad_pattern} (f), which effectively conveys the reflected signals toward the receiver. The relation between the input configurations and the output configuration can be clearly seen; however, it is not a straightforward procedure to obtain Fig. \ref{fig:rad_pattern} (c) from Figs. \ref{fig:rad_pattern} (a) and (b). The proposed CNN model will be trained to learn this complex relationship so that the final RIS configuration can be rapidly provided.


For an efficient training of the CNN model, the first stage of the proposed method aims to find horizontal and vertical configurations using an iterative algorithm that only requires M and N iterations, respectively. The ordering of this optimization (vertical/horizontal or horizontal/vertical) does not affect the results. Algorithm \ref{alg:ite} provides a pseudo-code for finding the most suitable horizontal configuration. At each iteration, a stripe of RIS elements is configured with a predefined state. The received signal power $P_{rx}$ is then calculated at the receiver. This iteration is then repeated with a different state for the same stripe. The algorithm returns the horizontal configuration, which provides the highest received power at the end of the iterations. A similar procedure is applied to find the vertical configuration. For the RIS with $M\times N$ elements and $P$ different phase states for each element, this algorithm requires $(M+N)\times P$ iterations. These two configurations are provided as inputs to the proposed CNN model that will generate a final configuration. The resulting configuration is expected to generate a radiation pattern with a maximum gain in the desired ($\vartheta_{0},\varphi_{0}$) direction.

\begin{algorithm}[h]
\caption{The G-IM for finding horizontal RIS configuration.} \label{alg:ite}
\begin{algorithmic}[1]
\renewcommand{\algorithmicrequire}{\textbf{Inputs:}}
\renewcommand{\algorithmicensure}{\textbf{Output:}}
\REQUIRE $M$, and $P$
\ENSURE  \textit{$S_h$} \textit{\# the states of the horizontal RIS elements}
\STATE initialize \textit{$S_h$} with $0^\circ$ phase shift
\STATE $P_{rx,max}$ $\gets - \infty$
\FOR {$i \in [1, M]$}
\FOR {$j \in [1, P]$ }
\STATE configure $i$th row with state $j$
\STATE calculate $P_{rx}$
\IF {($P_{rx} > P_{rx,max}$)}
\STATE update $i$th row of \textit{$S_h$} with state $j$
\STATE  $P_{rx,max}$ $\gets$ $P_{rx}$
\ENDIF
\ENDFOR
\ENDFOR
\RETURN \textit{$S_h$}
\end{algorithmic}
\end{algorithm}

Keras \cite{keras}, which is an open-source machine learning framework, is utilized to develop the CNN proposed model consisting of six 3D convolutional layers and two dropout layers. The 3D convolutional layers have 4, 16, 64, 8, 2, and 1
filters, respectively, with a fixed kernel size of (3 x 3 x 2), and the hyperbolic tangent ($\tanh$) activation function is used in each convolutional layer. In order to provide better generalization for the network and prevent over-fit data, the dropout layers are used with a rate of 0.2. Furthermore, the model parameters are determined using the adaptive moment estimation (ADAM) optimizer, and early stopping is used during the training phase to avoid over-fitting. For the early stopping function, a patience of 10 epochs is set, which checks the validation loss throughout training. If the validation loss converges to a level and stays there for 10 epochs, the training is stopped, and the weights from the final training run are utilized in the test. Table \ref{tab:CNN_arc} provides the design parameters for the proposed CNN model.
\begin{table}[]
\centering
\caption{The Proposed CNN Model Architecture}
\label{tab:CNN_arc}
\begin{tabular}{c|cc}
\hline
\textbf{Layer}       & \multicolumn{1}{c|}{\textbf{Output Shape}} & \multicolumn{1}{l}{\textbf{Filter Shape}} \\ \hline
Input                & \multicolumn{1}{c|}{40 x 40 x 2}           & -                                         \\
Conv3D\_1            & \multicolumn{1}{c|}{40 x 40 x 4}           & 3 x 3 x 2                                 \\
Conv3D\_2            & \multicolumn{1}{c|}{40 x 40 x 16}          & 3 x 3 x 2                                 \\
Conv3D\_3            & \multicolumn{1}{c|}{40 x 40 x 32}          & 3 x 3 x 4                                 \\
Dropout (0.2)             & \multicolumn{1}{c|}{40 x 40 x 32}          & -                                         \\
Conv3D\_4            & \multicolumn{1}{c|}{40 x 40 x 128}         & 5 x 5 x 8                                 \\
Conv3D\_5            & \multicolumn{1}{c|}{40 x 40 x 64}          & 5 x 5 x 8                                 \\
Conv3D\_6            & \multicolumn{1}{c|}{40 x 40 x 8}           & 3 x 3 x 4                                 \\
Dropout (0.2)             & \multicolumn{1}{c|}{40 x 40 x 8}           & -                                         \\
Conv3D\_7            & \multicolumn{1}{c|}{40 x 40 x 4}           & 3 x 3 x 2                                 \\
Conv3D\_8            & \multicolumn{1}{c|}{40 x 40 x 1}           & 3 x 3 x 2                                 \\
Output               & \multicolumn{1}{c|}{40 x 40}               & -                                         \\ \hline
Trainable Parameters & \multicolumn{2}{c}{2,496,665}                                                           
\end{tabular}
\end{table}
\section{Simulation Results}
In this section, we present simulation results for the RIS-assisted wireless communication system in Fig. \ref{fig:system_schema}. For training, validating, and testing the CNN model, we generate the dataset including horizontal, vertical, and reference RIS configurations for each angular position of the $R_x$. Horizontal and vertical RIS configurations are generated using the G-IM while the IM is utilized for the reference RIS configurations. We assume that the $R_x$ is located at the far-field of the RIS ($10$m away from RIS) with the azimuth and elevation angles in the range [$0^\circ$, $180^\circ$] and [$-60^\circ$, $60^\circ$], respectively. The dataset is generated using the resolution of $1^\circ$, therefore resulting in $181\times121$ different data points. Note that, for each data point, two images corresponding to vertical and horizontal RIS configurations, and the third image corresponding to the reference RIS configuration are generated. The dataset is divided into three groups, where the training phase uses $60\%$ of the data while the remaining data is split equally between the validation and test phases.
\begin{figure}[t]

\begin{minipage}[h]{0.48\linewidth}
    \centering
    \includegraphics[width=1\linewidth]{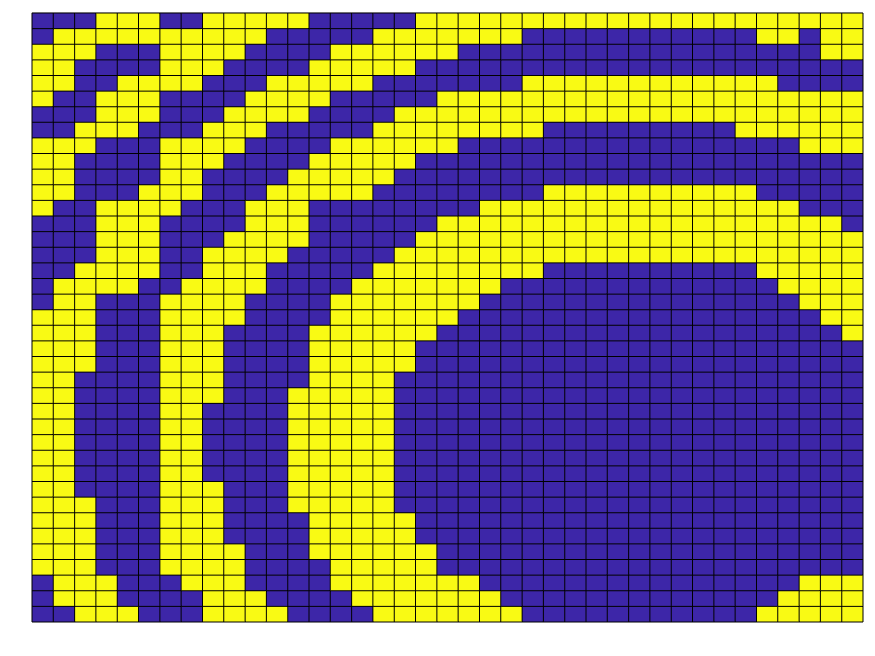}
    \subcaption{}
\end{minipage}
\begin{minipage}[h]{0.48\linewidth}
    \includegraphics[width=1\linewidth]{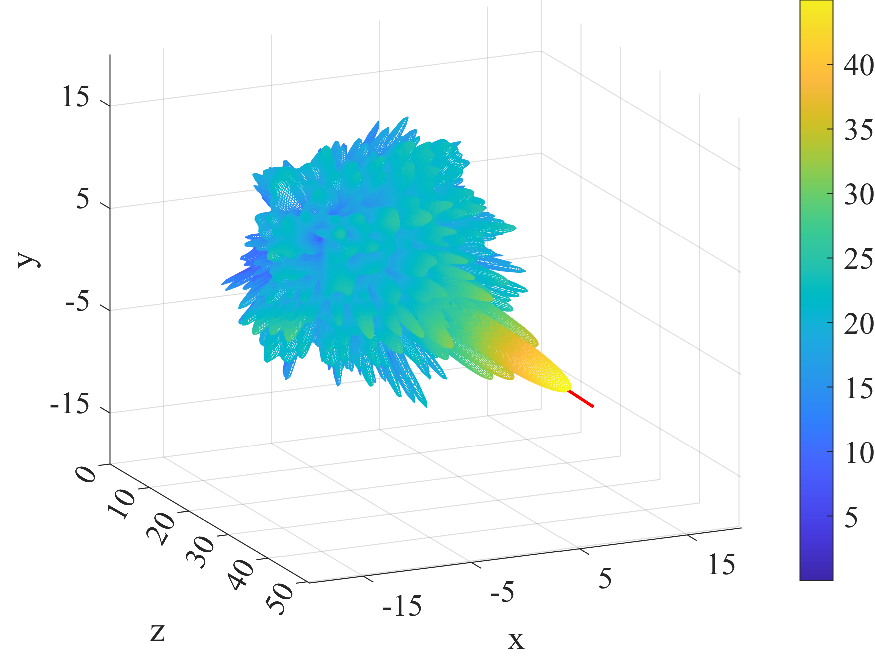}
    \subcaption{}
\end{minipage}
 \caption{(a) Estimated RIS configuration image of the proposed CNN model, and (b) the corresponding radiation pattern.}
\label{fig:est_config_and_pattern}  
\end{figure}
\begin{figure}[t]
    \centering
    \includegraphics[width=1\linewidth]{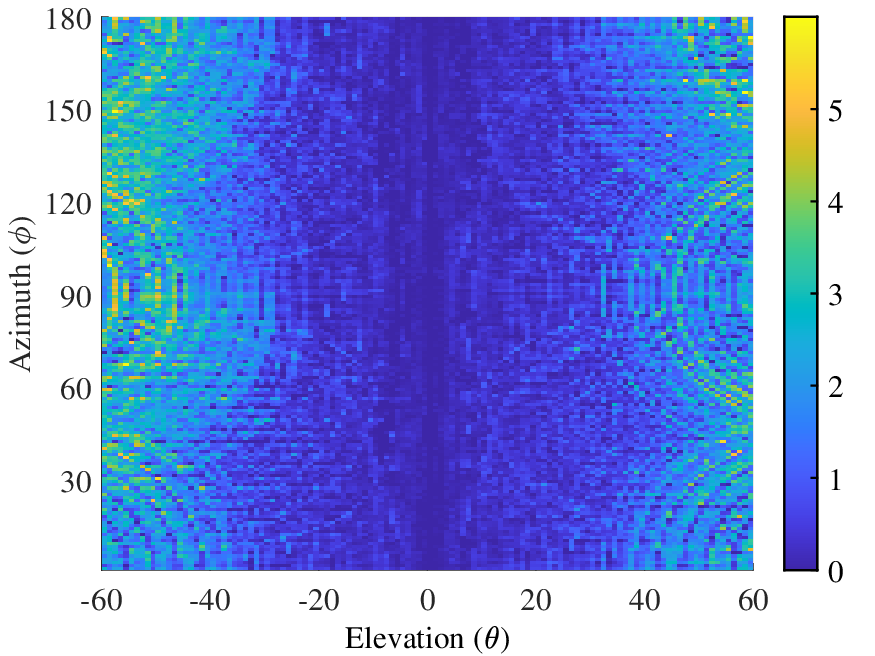}
    \caption{The received power differences between the IM and the CNN-G-IM for different angular positions of the receiver.}
    \label{fig:comparison_proposed}
\end{figure}

Fig. \ref{fig:est_config_and_pattern} shows an estimated configuration of the CNN-G-IM and its corresponding radiation pattern when the $R_x$ is located at the same position in Fig. \ref{fig:rad_pattern}. The results show that the CNN-G-IM provides a RIS configuration image similar to the reference image shown in Fig. \ref{fig:rad_pattern} (c). Although the generated configuration is not exactly the same as the reference, the radiation patterns are similar with a beam pointing in the desired direction (\ref{fig:rad_pattern} (c) and \ref{fig:est_config_and_pattern} (b)).

\begin{figure}[t]

\begin{minipage}[h]{1\linewidth}
    \centering
    \includegraphics[width=0.48\linewidth]{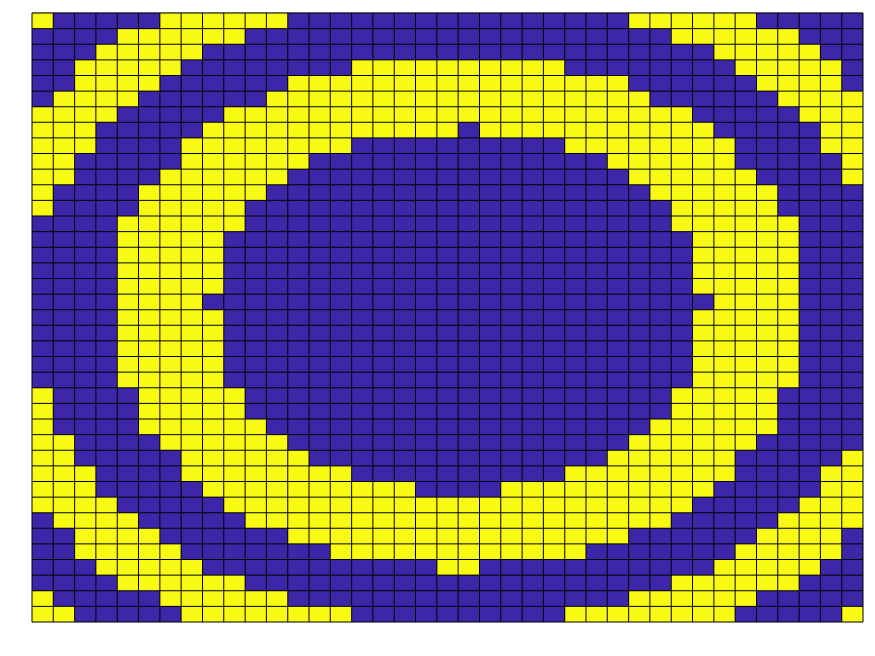}
    \includegraphics[width=0.48\linewidth]{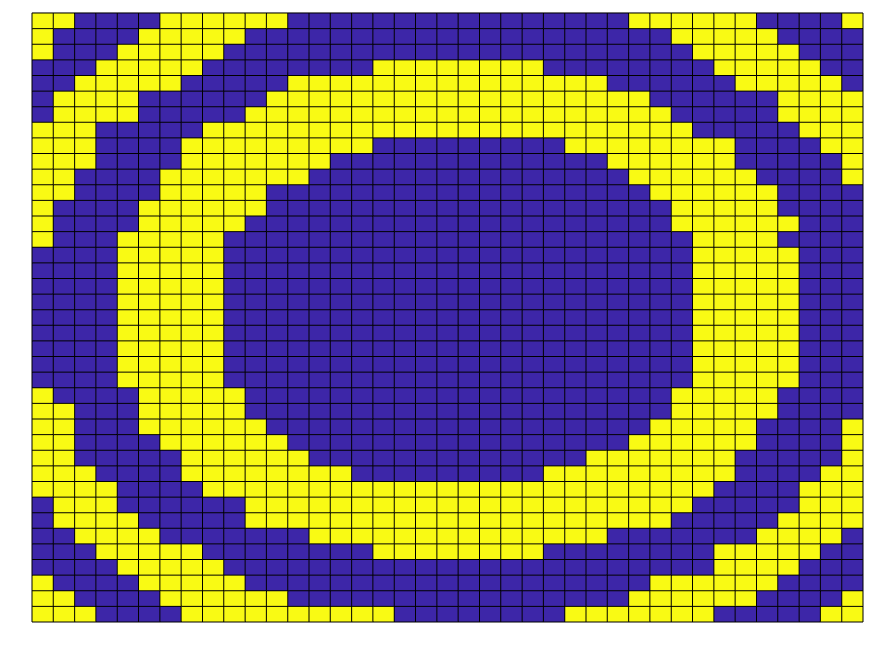}
    \subcaption{Angular position of the receiver: $\theta=0^\circ, \phi=0^\circ$}
\end{minipage}

\begin{minipage}[h]{1\linewidth}
    \centering
    \includegraphics[width=0.48\linewidth]{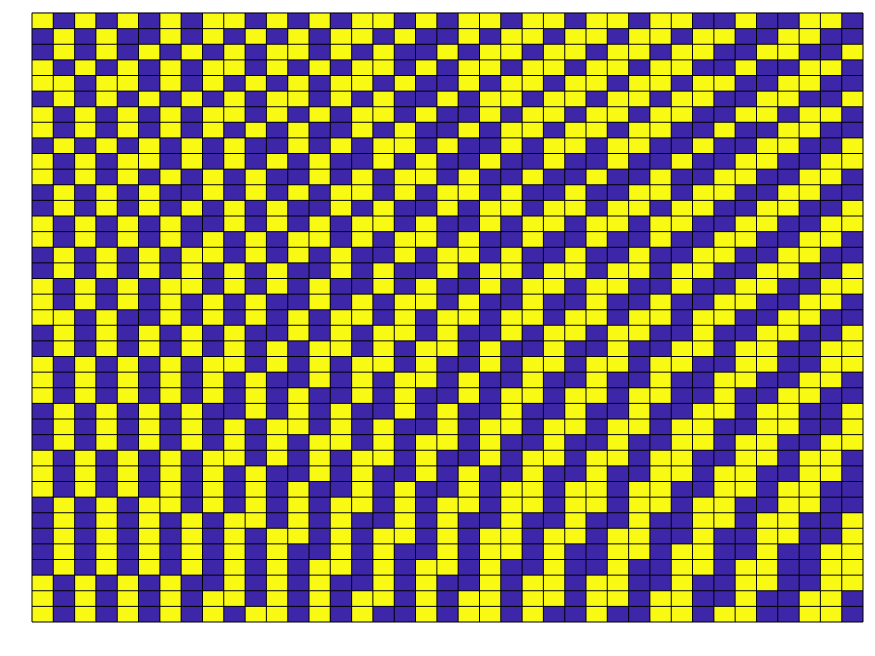}
    \includegraphics[width=0.48\linewidth]{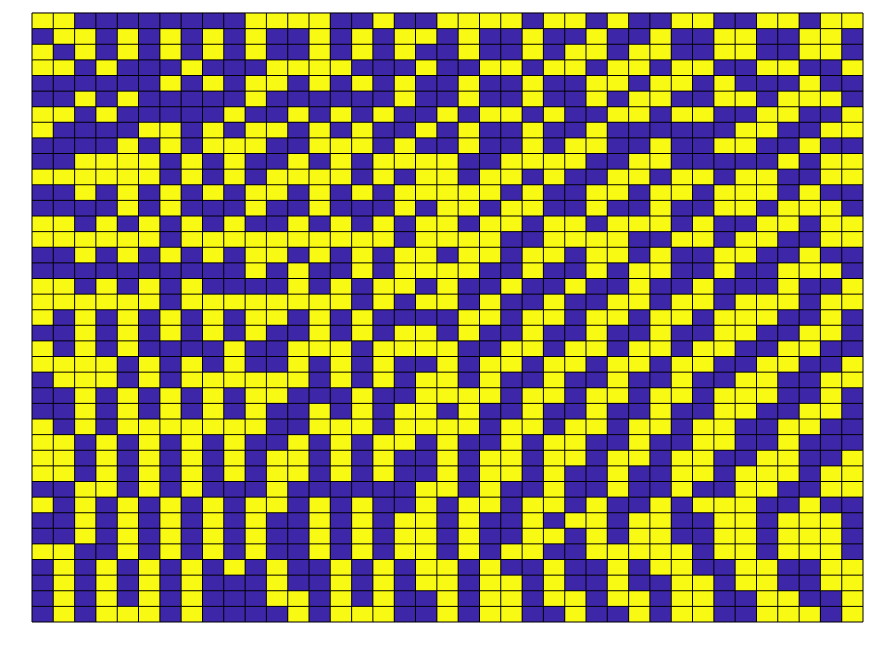}
    \subcaption{Angular position of the receiver: $\theta=60^\circ, \phi=300^\circ$}
\end{minipage}
\caption{The illustration of the reference RIS configuration image obtained by the IM (on the left) and the final RIS configuration image obtained by CNN-G-IM (on the right).}
\label{fig:config_ests} 
\end{figure}

Fig. \ref{fig:comparison_proposed} shows the difference of the received power when the radiation patterns are obtained using the RIS configurations of the CNN-G-IM and the IM for $181\times121$ data points. At each data point in the figure, the received powers of both IM and CNN-G-IM are recorded, and the received power of the CNN-G-IM is subtracted from the received power of the IM. The results demonstrate that the proposed CNN model provides a close performance to the IM, where the power difference values are mostly low. Specifically, the performance degradation occurs in the elevation angles corresponding to the edges in the figure, where the maximum power difference is 6dB. Furthermore, for an elevation interval of $-45^\circ$ to $45^\circ$, the performance difference is negligible. The reason is that the configurations required to form the peaks at the edges are more complex and challenging to learn compared to those for the middle data points in the figure. This is illustrated in Fig. \ref{fig:config_ests}. The configuration of Fig. \ref{fig:config_ests}(a) is learnable and can be estimated by the CNN as in Fig. \ref{fig:config_ests}(b). On the other hand, Figs. \ref{fig:config_ests}(c) and (d) show a sophisticated configuration and its coarse estimation by the CNN, respectively. 

\begin{figure}[t]
    \centering
    \includegraphics[width=0.97\linewidth]{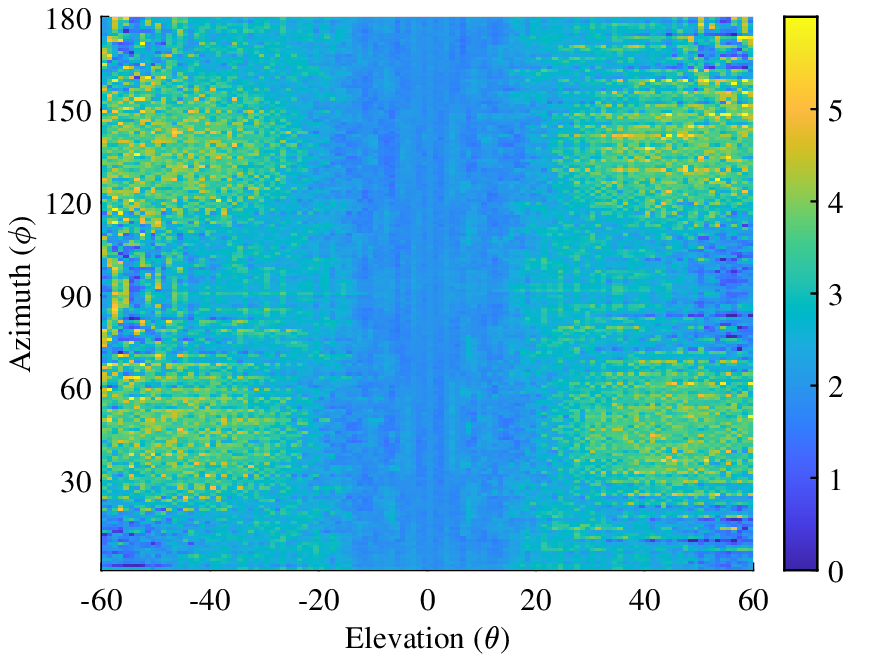}
    \caption{The power differences between the IM and the G-IM over the angular position of the receiver.}
    \label{fig:comparisonMNtoM+N}
\end{figure}

\begin{figure}[t]

\begin{minipage}[h]{0.48\linewidth}
    \centering
    \includegraphics[width=1\linewidth]{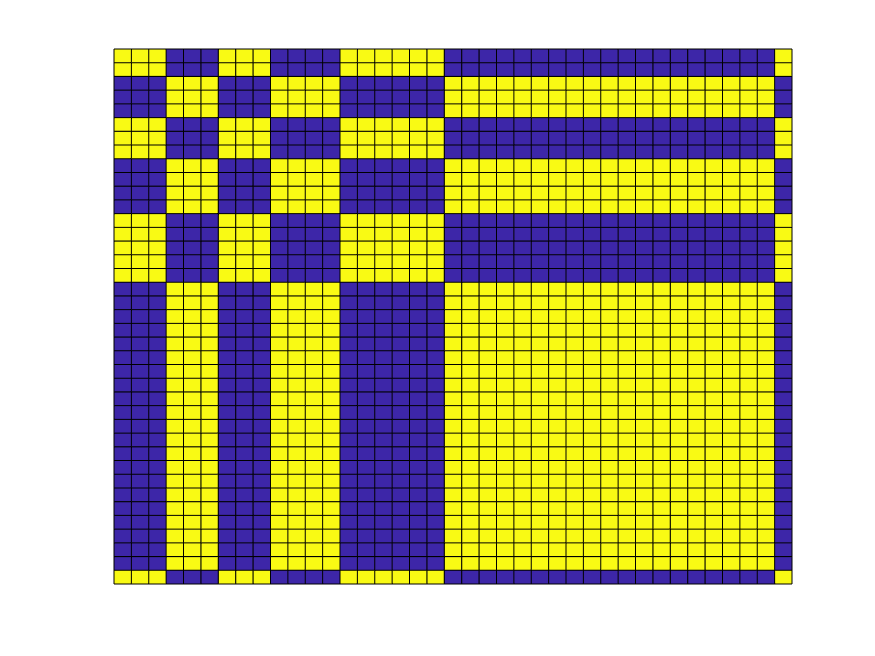}
    \subcaption{}
\end{minipage}
\begin{minipage}[h]{0.48\linewidth}
    \includegraphics[width=1\linewidth]{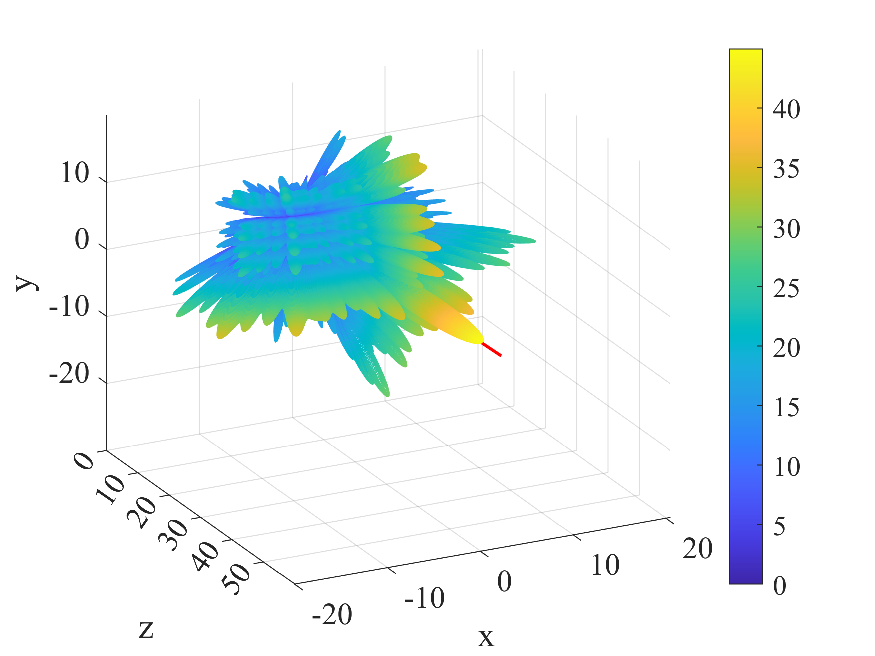}
    \subcaption{}
\end{minipage}
\caption{(a) The configuration found according to the algorithm given in \cite{MNtoM+N}  and (b) the radiation pattern of this configuration.}

\label{fig:h+w_rad}
    
\end{figure}

Note that the CNN model is utilized to enhance the performance of the G-IM. To visualize the performance improvement provided by the CNN model, Fig. \ref{fig:comparisonMNtoM+N} illustrates the difference of the received power when the G-IM in \cite{MNtoM+N} and the IM are employed. When the results in Fig. \ref{fig:comparison_proposed} and Fig. \ref{fig:comparisonMNtoM+N} are analyzed, we observe that the proposed CNN model significantly enhances the received power perforamnce of the G-IM since the values in Fig. \ref{fig:comparison_proposed} are significantly lower compared to the values in  Fig. \ref{fig:comparisonMNtoM+N}. According to \cite{MNtoM+N}, the resulting configuration and its radiation pattern for the same location in Fig. \ref{fig:est_config_and_pattern} are given in Fig. \ref{fig:h+w_rad}. As can be seen, although the maximum gain of the radiation pattern exists in the direction of the receiver, the efficiency loss is obvious due to the side lobe existing at horizontal and vertical beam planes. The model makes the resulting configuration to be smoother than the configuration generated by \cite{MNtoM+N} seen in Fig. \ref{fig:h+w_rad} (a), which suppresses the radiated power existing in the non-interested part of the horizontal and vertical beam planes that can be seen from Figure \ref{fig:h+w_rad} (b).



\section{Conclusion}
This paper presents a CNN-assisted group based iterative method for optimizing the RIS configuration. The proposed method reduces the number of iterations from $N\times M$ to $N+M$. The simulation results demonstrate that this significant reduction in the complexity is achieved with a slight degradation in the performance. As future work, different learning models can be developed to further improve the performance. The system model will also be extended to include multi-user scenarios in addition to  sensing applications such as the angle of arrival estimation. 

\section*{Acknowledgment}
This publication was made possible in parts by  NPRP13S-0130-200200  and by NPRP14C-0909-210008 from the Qatar National Research Fund (a member of The Qatar Foundation). The statements made herein are solely the responsibility of the authors.
We thank to StorAIge project that has received funding from the KDT Joint Undertaking (JU) under Grant Agreement No. 101007321. The JU receives support from the European Union’s Horizon 2020 research and innovation programme in France, Belgium, Czech Republic, Germany, Italy, Sweden, Switzerland, Türkiye, and National Authority TÜBİTAK with project ID 121N350.


\bibliographystyle{IEEEtran}
\bibliography{reference.bib}

\end{document}